\documentclass[aps, pra, superscriptaddress, twocolumn
]{revtex4-1}%
\usepackage{graphicx}
\usepackage{dcolumn}
\usepackage{bm}
\usepackage{hyperref}
\usepackage[normalem]{ulem}
\usepackage{color}
\usepackage{amsmath}
\usepackage{amsfonts}
\usepackage{amssymb}%

\begin{document}
	
	\title{A quantum optical model for the dynamics of high harmonic generation}

\author{\'{A}kos Gombk\"{o}t\H{o}}
\affiliation{Department of Theoretical Physics, University of Szeged, Tisza Lajos k\"{o}r%
\'{u}t 84, H-6720 Szeged, Hungary}

\author{Attila Czirj\'{a}k}
\affiliation{Department of Theoretical Physics, University of Szeged, Tisza Lajos k\"{o}r%
\'{u}t 84, H-6720 Szeged, Hungary}
\affiliation{ELI-ALPS, ELI-HU Non-profit Ltd., Dugonics t\'{e}r 13, H-6720 Szeged, Hungary}%

\author{S\'{a}ndor Varr\'{o}}
\affiliation{Wigner Research Centre for Physics, Konkoly-Thege
M. \'ut 29-33, H-1121 Budapest, Hungary}
\affiliation{ELI-ALPS, ELI-HU Non-profit Ltd., Dugonics t\'{e}r 13, H-6720 Szeged, Hungary}%

\author{P\'{e}ter F\"{o}ldi}
\affiliation{Department of Theoretical Physics, University of Szeged, Tisza Lajos k\"{o}r%
\'{u}t 84, H-6720 Szeged, Hungary}
\affiliation{ELI-ALPS, ELI-HU Non-profit Ltd., Dugonics t\'{e}r 13, H-6720 Szeged, Hungary}%
	
\begin{abstract}
We investigate a two-level atom in the field of a strong laser pulse. The resulting time-dependent polarization is the source of a radiation the frequency components of which are essentially harmonics of the driving field's carrier frequency. The time evolution of this secondary radiation is analyzed in terms of the expectation values of the photon number operators for a large number of electromagnetic modes that are initially in the vacuum state. Our method is based on a multimode version of the Jaynes-Cummings-Paul model and can be generalized to different radiating systems as well. We show, that after the exciting pulse, the final distribution of the photon numbers is close to the conventional (Fourier transform-based) power spectrum of the secondary radiation. The details of the high harmonic spectra are also analyzed, for many-cycle excitations a clear physical interpretation is given in terms of the Floquet quasi-energies. A first step towards the determination of the photon statistics of the HHG modes reveals states with slightly super-Poissonian distribution.
\end{abstract}


	\maketitle

\section{Introduction}
High harmonic generation (HHG) \cite{F88,LB93,W93,MKG93} is a strongly nonlinear effect that is observed in several state of the art experiments, using gaseous targets (see e.g. \cite{WSG08} for a review), plasma surfaces \cite{TG09} and -- more recently -- solid state samples as well \cite{G10}. One of the most important applications is the generation of attosecond pulses \cite{FT92,KI09}, which can monitor or induce physical processes on an experimentally unprecedented time scale. Therefore, deep understanding of the physical mechanisms underlying the phenomenon of HHG is of crucial importance.

Although HHG is an inherently high-field effect, due to the low efficiency of the process, the intensities of the generated harmonics are by orders of magnitude lower than that of the exciting field. Therefore, the usual assumption \cite{C93b,L94} that the exciting field (with high photon numbers) not necessarily needs to be quantized, can be verified, but the same does not hold for the weak secondary radiation. Motivated by this, in the following we introduce a model where a quantum system interacts with a strong, classical electromagnetic field as well as with quantized radiation modes that are initially in the vacuum state.

The main features of the gas HHG spectra are well described by the so-called "three-step" model \cite{C93b}, consisting of the emission of the single active electron, its motion in the laser field and recombination with the parent ion. The amount of energy the electron gained during this process is assumed to be transferred into high harmonic radiation. In this picture the continuum plays a substantial role, and the emerging exponential integrals can be performed both in the adiabatic \cite{L94} and in the nonadiabatic \cite{Sa04} cases by means of saddle point approximation.

On the other hand, the appearance of the HHG itself does not require the presence of the continuum. As studies with driven two-level systems (having only bound states) show \cite{KS94,BF96,GGK97,PFM01}, the qualitative properties of the HHG spectra can be calculated analytically using traditional quantum optical notions like e.g., dressed states. An appropriately generalized version of the Jaynes-Cummings-Paul model has also been used to describe high-field, multiphoton processes \cite{SB81,D92,FO03}.  Although a strong analogy can be drawn between the dynamics of a driven two-level system and the three-step model \cite{FR03}, for the realistic description of HHG in gas samples (with the obvious involvement of continuum states), the latter one became more widely used. For solid state systems \cite{G10}, on the other hand, only bound states get populated, and even the two-level approximation can be valid for quantum wells in semiconductor heterostructures \cite{H94}.
\bigskip

Motivated by this, we use a two-level system as a model for the radiating quantum system (so that the resulting approach will be similar to that used in driven spin-boson models \cite{OIL13} or for the description of resonance fluorescence \cite{MS91}). Following the tradition, we keep using the term "atom", although as we have seen above, solid state systems can be described by this model more appropriately. The simplicity of the two-level system helps understanding the dynamics of the HHG, and -- as we shall see -- certain important aspects of our approach can be easily generalized to more complex high harmonic sources as well.  Let us emphasize that although many important results were obtained using a two-level system as a model -- even in the context of HHG (see e.g.~\cite{KS94,BF96,GGK97,PFM01}) -- we are not aware of any work focusing on the quantum optical description of the high harmonic modes. Additionally, the focus is usually on the spectrum of the secondary radiation, while our model allows monitoring the time evolution of the photon numbers corresponding to different modes of the emitted electromagnetic field.

The current paper is organized as follows. In Sec.~\ref{modelsec} we introduce the model. The time evolution of the photon number expectation values is presented in Sec.~\ref{expdynsec}, and the corresponding HHG spectra are analyzed in Sec.~\ref{dynsec}. We discuss the photon statistics of the high harmonic modes in Sec.~\ref{schsec}, and draw the conclusions in Sec.~\ref{conclusionsec}.

\section{Model}
\label{modelsec}
Let us consider the following Hamiltonian:
\begin{equation}
H(t)=H_{a}+H_{m}+H_{am}+H_{ex}(t),
\label{Ham}
\end{equation}
where the first term describes a two-level atom and the second one corresponds to quantized radiation modes:
\begin{equation}
H_{a}=\hbar\frac{\omega_{0}}{2}\sigma_z, \ \ H_{m}=\sum_n \hbar \tilde{\omega}_n a_n^{\dagger}a_n,
\label{H0}
\end{equation}
while their interaction [without rotating wave approximation (RWA)] can be written in the following form:
\begin{equation}
H_{am}=\sum_n \hbar \frac{\Omega_n}{2}(a_n+a_n^{\dagger})\sigma_x,
\label{Hint}
\end{equation}
where $\sigma_x$ and $\sigma_z$ denote the usual Pauli matrices. Note that $H_{am}=-D E,$ where the atomic dipole moment operator is given by $D=d\sigma_x$ (for the sake of simplicity, the matrix element $d$ is chosen to be real), and the quantized electric field is a sum of operators $E_n=\sqrt{\frac{\hbar \tilde{\omega}_n}{\epsilon_0 V}}(a_n+{a_n}^{\dagger})$ (with $V$ denoting the quantization volume). This means that $\Omega_n=2d\sqrt{\frac{\hbar \tilde{\omega}_n}{\epsilon_0 V}}.$
Additionally, we assume a strong, pulsed excitation that is described by a classical, time-dependent electric field:
\begin{equation}
H_{ex}(t)= -DE(t)=-d\sigma_x E(t)=-\hbar\frac{\Omega(t)}{2} \sigma_x,
\end{equation}
where
\begin{equation}
E(t)=E_0\sin^2\left(\pi \frac{t}{\tau}\right) \cos(\nu t),
\label{field}
\end{equation}
if $0<t<\tau,$ and $E(t)=0$ otherwise. $T=2\pi/\nu$ will be used to denote the cycle time of the carrier wave. The initial state of the system will be assumed to be a tensorial product state,
\begin{equation}
|\Psi(t=0)\rangle_{am}=|\phi\rangle_{m0}|\psi\rangle_{a0},
\end{equation}
where the modes at $t=0$ are in their vacuum state
\begin{equation}
|\phi\rangle_{m0}=|0,0,\ldots 0\rangle.
\label{vac}
\end{equation}
Note that for semiconductor quantum dots  -- to which our model is most directly applicable -- dipole moment matrix elements $d$ range between 1 to 100 Debye, depending mainly on the size of the dots and their constituent materials. For practically resonant near infrared excitations, the condition $E_0 d=\hbar\nu$ means peak field amplitudes $E_0$ of the order of GV/m, which can be realized relatively easily using currently available Ti:Sapphire technology. The corresponding pulse durations are in the femtosecond range.
\bigskip

The dynamics induced by the Hamiltonian (\ref{Ham}) practically cannot be solved without approximations. Let us summarize the general physical and technical properties of the model, which can serve as basis of the approximative methods. Using Eq.~(\ref{vac}) and assuming that the atomic system is initially in its ground state, if there is no external driving (i.e., $|\psi\rangle_{a0}=|g\rangle$, $E_0=0$), there is essentially no dynamics. On the other hand, with $|\psi\rangle_{a0}=|e\rangle,$ there are interfering vacuum Rabi oscillations that can serve as a simple model for spontaneous emission: The excitation of the atomic system is distributed among the degrees of freedom of the radiation modes. This is the point where the question which modes we take into account becomes important. In principle, the sum in Eq.~(\ref{H0}) is infinite. However, when focusing on the process of HHG, only a finite number of harmonics (a few times ten) appear, so only modes with frequencies roughly in the range of $0$--$100$ $\nu$ play relevant role. Practically (from the viewpoint of numerical feasibility), a few thousand modes can be taken into account in this interval. As a consequence, the initially excited atomic state will not monotonically decay: when oscillations with the finite number of Rabi frequencies $\Omega_n$ rephase, we would observe a revival process, that does not appear in free space. Technically, this allows us the determination of the time scale on which our model with a finite number of modes describes the process appropriately: Since for usual excitations, spontaneous emission plays negligible role (the lifetimes of the atomic levels can be orders of magnitude longer than the duration of the HHG process), our theoretical description is valid until the quantized modes do not cause significant atomic decay. This limit can be identified using our model based on expectation values (see the next section), and the parameters found to be valid for that calculation will be used later on to calculate e.g., photon statistics.

As a final, general note, let us mention that neither specific mode functions, nor the density of modes were taken explicitly into account. In an actual experiment, these issues must be relevant, but it is not inconsistent to neglect them on the level of the model of two-level atoms. Additionally, although a complete description of the HHG signal emerging from a given sample requires taking propagation effects also into account (see e.g.~\cite{BBT14}), here we focus on the single-atom response.

\section{Time evolution of the photon number expectation values}
\label{expdynsec}
Our first approach is based on the Heisenberg equations of motion for both the atomic and electromagnetic field operators. As one can check easily, one cannot obtain an exact closed set of dynamical equations, since the time derivatives of two-operator products involve three-operator products, whose time derivatives contain terms being the products of four operators, etc. Because of this hierarchy, we have to restrict ourselves to expectation values and -- at some point -- introduce a factorization that is based on physical considerations.

The simplest Heisenberg equations of motion read:
\begin{eqnarray}
\frac{d}{dt}\sigma_x&=&\omega_0 \sigma_y, \label{der1}\\
\frac{d}{dt}\sigma_y&=&-\omega_0 \sigma_x+\sigma_z \left[\Omega(t)+\sum_n \Omega_n \left(a_n+a_n^{\dagger}\right)\right], \label{der2}\\
\frac{d}{dt}\sigma_z&=&-\sigma_y \left[ \Omega(t) + \sum_n \Omega_n (a_n+a_n^{\dagger})\right], \label{der3}\\
\frac{d}{dt}a_n&=&-i\left(\tilde{\omega}_n a_n+\frac{\Omega_n}{2}\sigma_x\right),\label{der4} \\
\frac{d}{dt}N_n&=&\frac{d}{dt} a_n^{\dagger}a_n= i\frac{\Omega_n}{2}\left(a_n-a_n^{\dagger}\right)\sigma_x \label{der5}.
\label{Heis}
\end{eqnarray}
One can assume already at this point that the expectation values of the products of atomic an field operators factorize (e.g., $\langle\sigma_z a_n\rangle=\langle\sigma_z\rangle \langle a_n\rangle$), but in this case the influence of the photon number operators on the atomic system's time evolution -- which should be weak -- is exactly zero. Additionally, since the photon annihilation operators' dynamics are governed by a "classical" quantity (the expectation value of the dipole moment), the resulting photon statistics is necessarily Poissonian (i.e., the high harmonic modes are in coherent states).
\begin{figure}[htb]
\includegraphics[width=8.5cm]{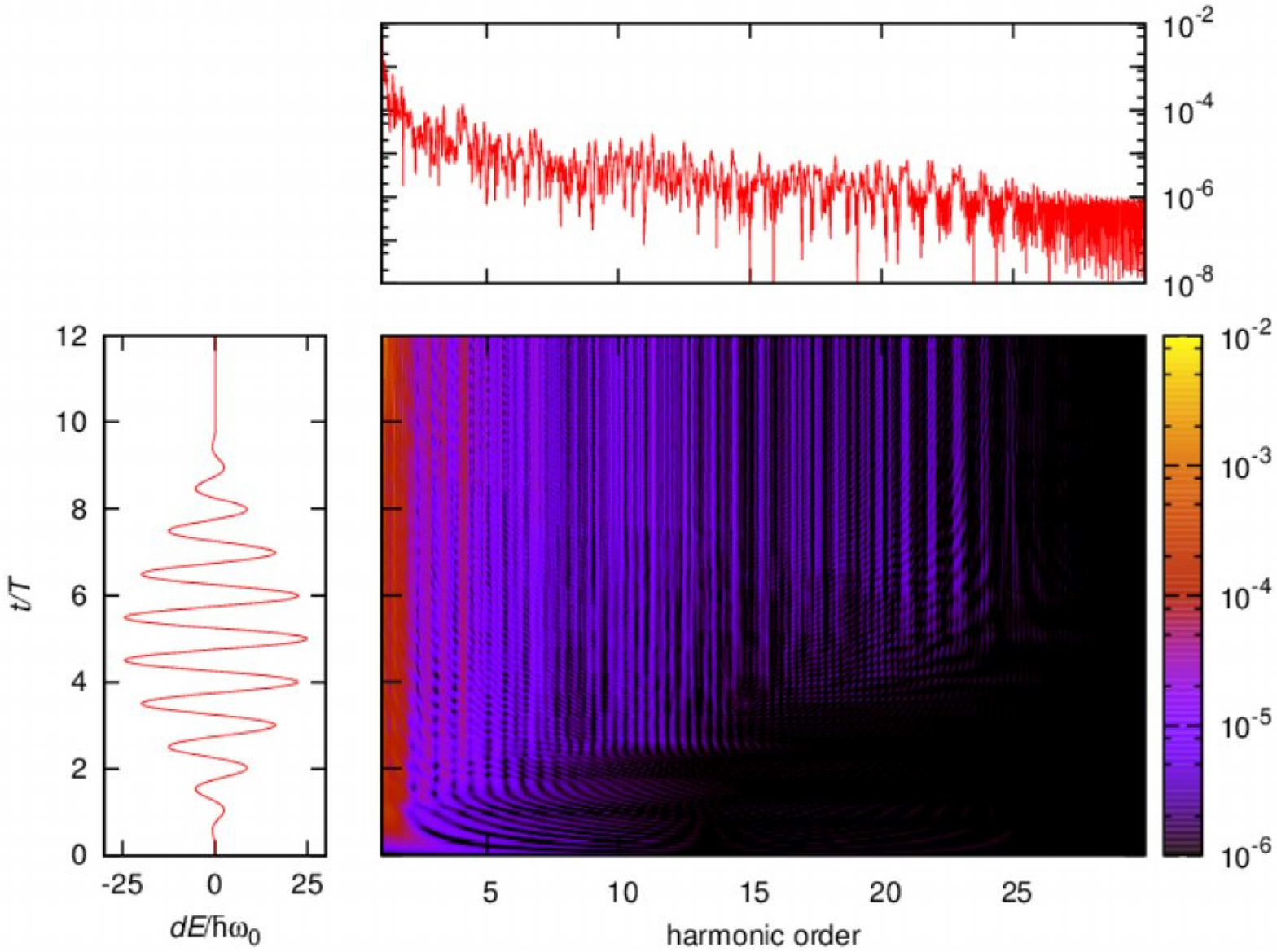}
\caption{Central panel: the time evolution of the expectation values of the photon number operators $\langle N_n \rangle$ as a function of the frequency of the modes $\tilde{\omega}_n$ (measured in units of $\nu$). The time dependence of the resonant exciting pulse ($\nu=\omega_0$) can be seen in the left panel, while the final ($t=12$ $T$) distribution of the photon number operator expectation values is shown on the top.}
\label{nexpfig1}%
\end{figure}

The next level of equations of motion can be obtained by assuming that the dynamical variables are the following expectation values of Hermitian operators that appear in Eqs.~(\ref{der1})-(\ref{der5}):
\begin{equation}
U^{+}_n=\left\langle \sigma_x (a_n + a_n^{\dagger})\right\rangle, \ \
U^{-}_n=i\left\langle \sigma_x (a_n - a_n^{\dagger})\right\rangle
\end{equation}
and similarly
\begin{eqnarray}
V^{\pm}_n=(i)^{(1\mp1)/2}\left\langle \sigma_y (a_n\pm a_n^{\dagger}) \right\rangle,\\
W^{\pm}_n=(i)^{(1\mp1)/2}\left\langle \sigma_z (a_n\pm a_n^{\dagger}) \right\rangle,
\end{eqnarray}
which are analogous to the "phonon assisted" density matrix elements that appear e.g., in semiconductor Bloch-equations \cite{HK04}. Using $U=\langle\sigma_x\rangle,$  $V=\langle\sigma_y\rangle,$ $W=\langle\sigma_z\rangle$ and $\langle N_n \rangle,$ a closed set of dynamical equations can be obtained (see the Appendix), provided one neglects cross correlations of different modes. (More precisely, with $A$ being either $U,$ $V$ or $W,$ we set $\langle A b_n c_m\rangle=0,$ if i) $n\neq m,$ and ii) the mode operators $b_n$ and $c_n$ are both creation or both annihilation operators. We will verify this assumption in Sec.~\ref{schsec}.)

\begin{figure}[htb]
\includegraphics[width=8.5cm]{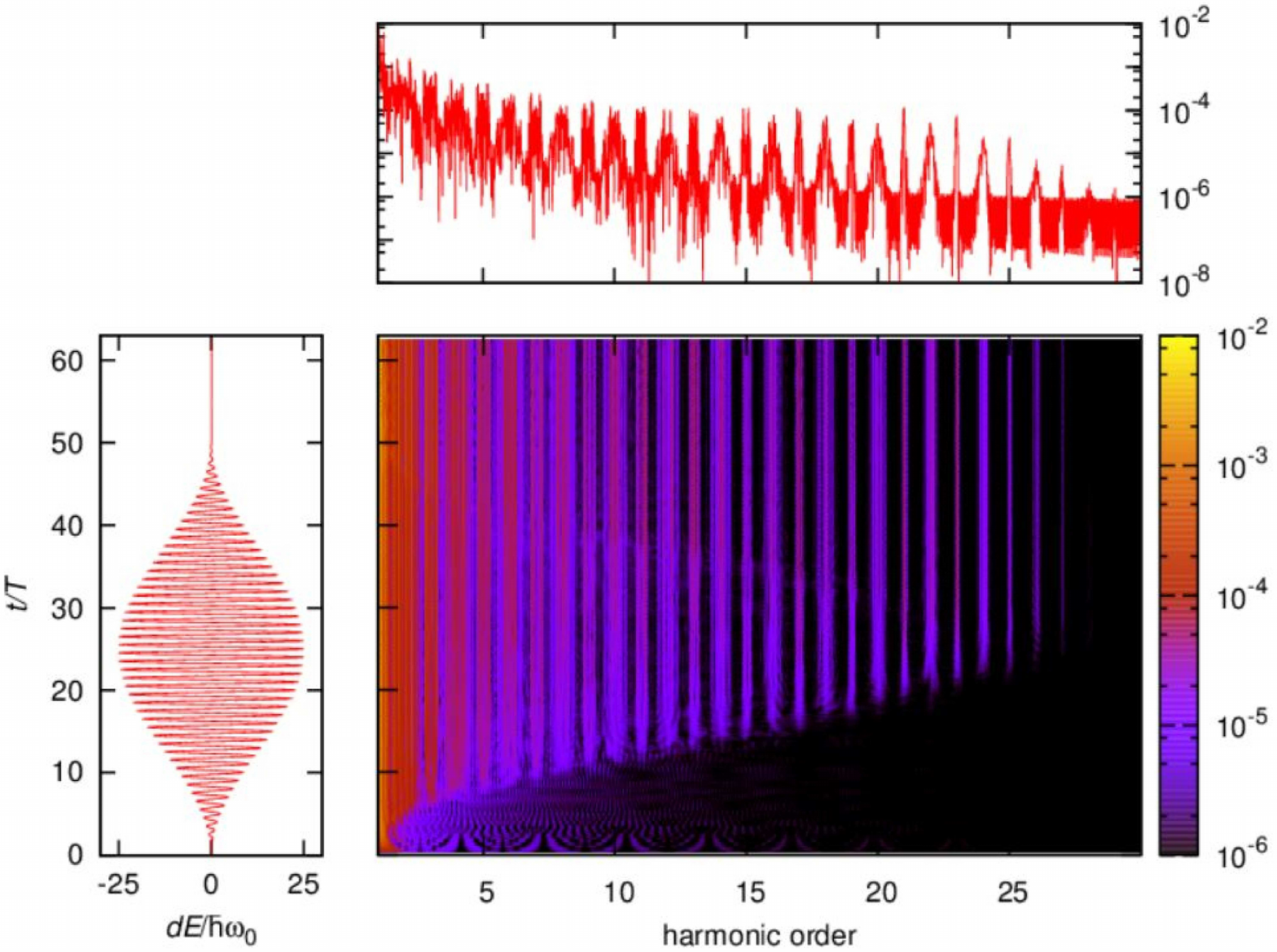}
\caption{The same as Fig.~\ref{nexpfig1}, but for a longer, resonant exciting pulse (compare the left panels).}%
\label{nexpfig2}%
\end{figure}

\begin{figure}[htb]
\includegraphics[width=8.5cm]{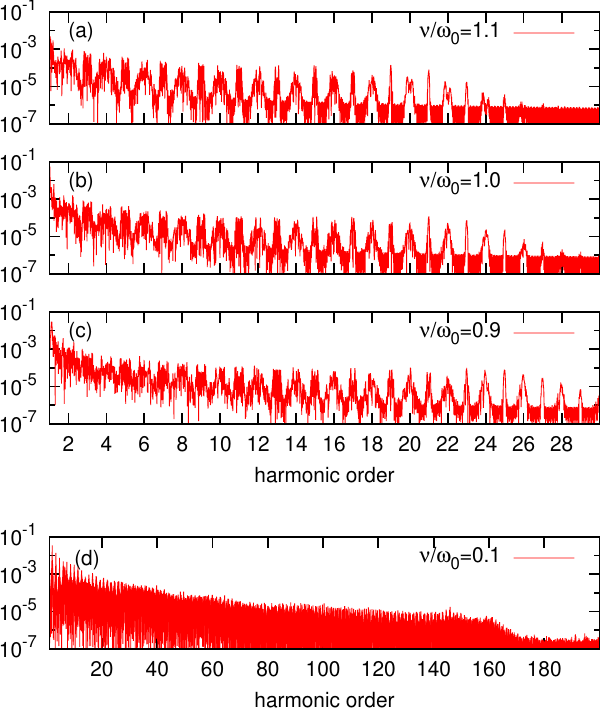}
\caption{The final ($t>\tau$) distribution of the photon number expectation vales for excitations with different frequencies (see the legend). The time evolution of the exciting pulses are the same in all subfigures as the one shown in the left panel of Fig.~\ref{nexpfig2}, provided we measure time in units of $T.$ (Without this scaling, the duration of the pulse corresponding to the top subfigure is the shortest.)}
\label{resnonresfig}%
\end{figure}

Technically, this approach, with $N$ modes being taken into account, means following the dynamics of $7N+3$ real variables. According to our experience, for most of our results, the frequency interval $[0, 30\nu]$ with $N=3000$ together with the realistic assumption of $\Omega_n/\omega_0=0.001\sqrt{\tilde{\omega}_n/\omega_0}$ satisfy the requirement mentioned at the end of the previous section: even an exciting pulse as long as few hundred optical cycles is much shorter than the time scale on which the quantized modes observably modify the atom's dynamics.
This means that the number of modes that we take into account has negligible influence on the time evolution of the atom, it is mainly the resolution of the spectra that is determined by the value of $N.$ By requiring 100 frequency values (points on the graphs) in an interval of "length" $\nu,$ the value of $N=3000$ means covering the interval of $[0, 30\nu]$ sufficiently densely. For most of the cases we considered, this is appropriate, since no harmonics of higher order than 30 appear. However, for the strongly detuned excitation shown in Fig.~\ref{resnonresfig}(d), there are roughly ten times more observable harmonics and the value of $N$ had to be increased correspondingly.

As an indication of the correctness of the factorization described above, let us mention that without excitation ($E_0=0$), the initial conditions $\langle N_i\rangle=n\delta_{i0}$ (where $\tilde{\omega}_0=\omega_0$) lead to Rabi flopping between mode $0$ and the atomic system with a frequency very close to $\Omega_0\sqrt{n+1}.$

\smallskip

The time evolution of the photon number expectation values is shown in Figs.~\ref{nexpfig1} and \ref{nexpfig2}. As we can see, the longer the exciting pulse is, the more pronounced peaks can be observed. As one can expect, these peaks appear around integer multiples of the carrier frequency of the exciting pulse. For long enough pulses (see Fig.~\ref{nexpfig2}) and higher harmonic orders, a significant, qualitative difference can be perceived between the consecutive peaks: the ones at odd multiples of $\nu$ are considerably broader than those corresponding to even harmonics. We will return to this point in the next section.

The time instants when the peaks corresponding to the various harmonic orders appear are related to the presence of sufficiently strong excitation: at the beginning of the pulse, when the envelope of the exciting electric field is far from its maximum, lower order harmonics appear. The higher order ones become visible only around the maximum of the pulse. Note that -- according to our calculations (not shown here) -- for a simply sinusoidally oscillating excitation, the transient time interval is practically the same for all harmonics, in other words, the corresponding peaks become observable almost at the same time.

Fig.~\ref{resnonresfig} shows the effect of detuning on the final ($t>\tau$) distribution of the photon number expectation values. As we can see, not only the internal structure and height of the peaks, but even the number of the observable harmonics strongly depends on the frequency of the excitation. As we shall see in the next section, all these properties can be understood in terms of the Floquet quasi-energies and the corresponding states.

Note that the features of typical HHG spectra can be recognized in Fig.~\ref{resnonresfig}: the heights of the peaks corresponding to low order harmonics is decreasing fast, then we see a "plateau" with comparable peak heights and finally a cutoff, i.e., the disappearance of the pronounced peaks. However, focusing on the details, and comparing these spectra to the high harmonic emission spectra computed for real noble gas atoms (i.e. the real single atom response) \cite{IKS15} shows certain qualitative differences.
Close to resonance, the peaks of the two-level atom's spectrum emerge from a smooth background, which is mostly missing from the real single atom response. This is more explicit around the cutoff region, which means simply the disappearance of the pronounced peaks in the case of the two-level atom, while it is a smooth but substantial drop to much lower spectral amplitude in the real single atom response. Although the cutoff in the strongly detuned case [Fig.~\ref{resnonresfig}(d)] seems more similar to this real single atom spectrum, this is only apparent: the cutoff consists of densely placed peaks in the two-level case. These differences are due to the different physical mechanisms being responsible for the process of HHG for a real atom and a two-level one.

As we shall see in the next section, the final photon number distributions shown in Fig.~\ref{resnonresfig} are very close to the HHG spectra that can be obtained in the usual way, i.e., by calculating the power spectrum of the second time derivative of the dipole moment \cite{B09}. The dependence of these power spectra on the excitation parameters were studied extensively, e.g., analytic formulas for the plateau and the cutoff frequency were obtained in Refs.~\cite{KS94,BF96,GGK97}. Our results show good agreement with these analytic estimations. Therefore, instead of investigating these details further, we turn to the interpretation of the spectra and the analysis of the corresponding photon statistics.

\section{Properties of the HHG spectra}
\label{dynsec}

\begin{figure}[htb]
\includegraphics[width=8.5cm]{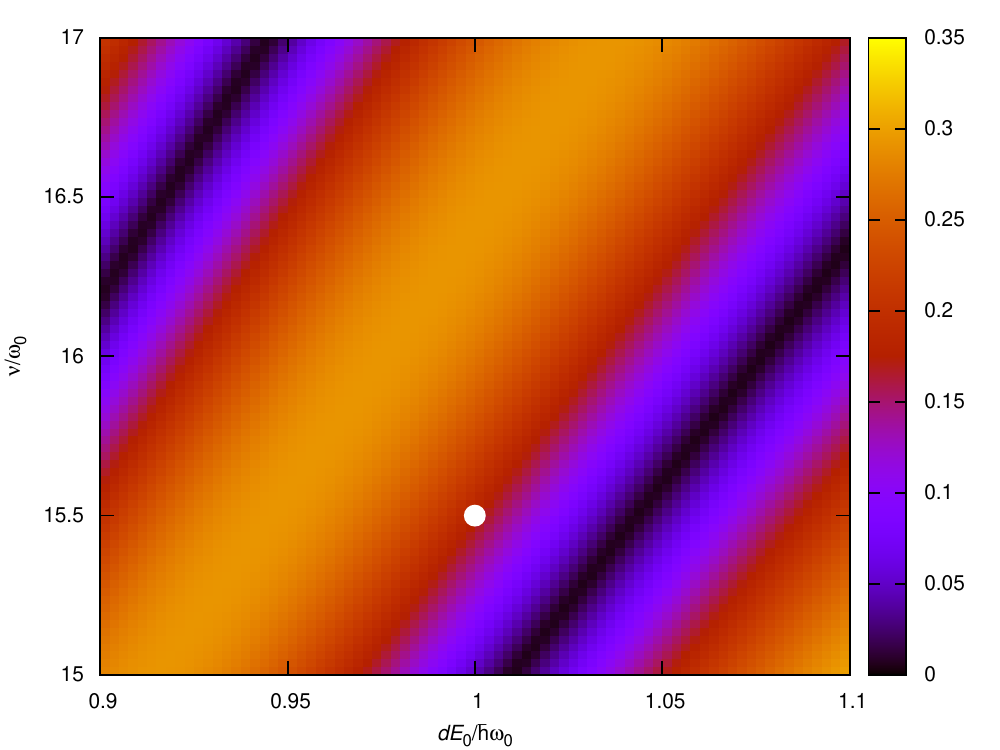}
\caption{The smallest difference between two nonequivalent Floquet quasi-energies ($\delta \epsilon=|\epsilon_1-\epsilon_2|$, measured in units of $\nu$) as a function of the amplitude and frequency of the exciting field. Note that in this case, by construction, the excitation is assumed to be monochromatic with an amplitude of $E_0.$ The white circle denotes parameters corresponding to Fig.~\ref{spectra2fig}.}%
\label{Floquetfig}%
\end{figure}

As we can see in Fig.~\ref{resnonresfig}, HHG spectra show an alternating sequence of narrow and broader peaks. For long pulses, this effect can be understood via the Floquet analysis \cite{F883} of the problem. Note that although this idea itself can be useful for more complex radiating systems as well (with obviously increasing computational costs of performing the analysis, see e.g.~Ref.~\cite{PS89} for early results related to atomic hydrogen), this section is specific to two-level systems. (In other words, e.g., neither the internal structure nor the heights of the HHG peaks can be transferred to different systems.)

In order to see the predictions of Floquet's theory, let us consider a sinusoidal exciting field
\begin{equation}
E'(t)=E_0\sin(\nu t),
\end{equation}
which results in a periodic Hamiltonian
\begin{equation}
H'(t+T)=H_{a}-DE'(t+T)=H'(t),
\label{Hamprime}
\end{equation}
where the weak corrections $H_m$ and $H_{am}$ are neglected, and $T=2\pi/\nu.$ According to the general theorem \cite{F883,S65}, $H'(t)$ has "time dependent eigenstates"
\begin{equation}
\left|\phi_k (t)\right\rangle=e^{-i\epsilon_k t}\sum_n c_n^k e^{-i n \nu t} |\varphi\rangle_n^k,
\label{tdeigen}
\end{equation}
where $\epsilon_k$ are called Floquet quasi-energies (now written in units of frequency).
There are only two of the states above that are not equivalent (and these states are orthogonal). (The formulae $\epsilon_k\rightarrow \tilde{\epsilon}_k = \epsilon_k+m\nu,$ $c_n^k\rightarrow \tilde{c}_{n}^k=c_{n-m}^k$ define the equivalence, resulting in physically undistinguishable states $\left|\phi_k (t)\right\rangle$ and $\left|\tilde{\phi}_k (t)\right\rangle.$) Let us use indices 1 and 2 for the two nonequivalent states the quasi-energies of which have the smallest magnitude. This means that $\delta \epsilon=|\epsilon_1-\epsilon_2|<\nu.$  The dependence of $\delta \epsilon$ on the amplitude $E_0$ and the detuning $\Delta=\omega_0-\nu$ can be seen in Fig.~\ref{Floquetfig}. (Note that if we applied RWA, $\delta \epsilon,$ which is essentially the Rabi frequency, would scale linearly with $E_0$ and change as $\delta \epsilon (\Delta)=\sqrt{(\delta\epsilon(0))^2+\Delta^2}.$)

Let us assume that the state of the atomic system at $t=0$ can be expanded as
\begin{equation}
|\Psi(0)\rangle=\alpha |\phi_1(0)\rangle+ \beta |\phi_2(0)\rangle.
\end{equation}
The time dependence of the expectation value of the dipole moment operator reads
\begin{eqnarray}
\label{Fleq}
\langle D \rangle (t)&=&\langle \Psi(t)|D|\Psi(t)\rangle= \\
&=&|\alpha|^2 \langle \phi_1(t)|D|\phi_1(t)\rangle +
|\beta|^2 \langle\phi_2(t)|D|\phi_2(t)\rangle \nonumber \\
&+&\alpha\beta^* \langle\phi_2(t)|D|\phi_1(t)\rangle
+\alpha^*\beta \langle\phi_1(t)|D|\phi_2(t)\rangle \nonumber,
\end{eqnarray}
and -- as we can check easily -- contains frequency components $n\nu$ (the first two terms on the rhs of the equation above) and $n\nu\pm \delta\epsilon$ (the last two terms).
Additionally, the symmetry of the time periodic Hamiltonian (\ref{Hamprime}) implies \cite{SLD01} that $c_n^1$ is nonzero only for odd $n,$ while the frequency components of $|\phi_2(t)\rangle$ can be written as $\epsilon_2 +m\nu,$ with $m$ being even. Combining these, we obtain that for infinite, periodic excitation, the Fourier spectrum of $\langle D \rangle (t)$ contains discrete peaks that are situated at $m\nu$ for odd values of $m,$ and at $m\nu\pm\delta\epsilon$ when $m$ is even. In other words, this Fourier spectrum -- as well as that of $d^2 \langle D \rangle/dt^2$ -- is an alternating sequence of single and double peaks, which is in qualitative agreement with Figs.~\ref{nexpfig2} and \ref{resnonresfig}.
Let us emphasize that the qualitative difference between the HHG peaks corresponding to odd and even harmonics has symmetry related origin, similarly to the case of HHG spectra in gas samples \cite{F88}.

An additional remarkable point to be noticed is the appearance of the expansion coefficients $\alpha$ and $\beta$ in Eq.~(\ref{Fleq}). This fact directly shows that the state of the atomic system at $t=0$ plays an important role in determining the relative heights of the harmonic peaks. Specifically, when $\alpha=0$ (or $\beta=0$), the peaks corresponding to even (odd) harmonics are completely absent. Note that similar consequences of the atomic coherence have been pointed out e.g.~in Refs.~\cite{GK95,WS96}.

\begin{figure}[htp]
\includegraphics[width=8.5cm]{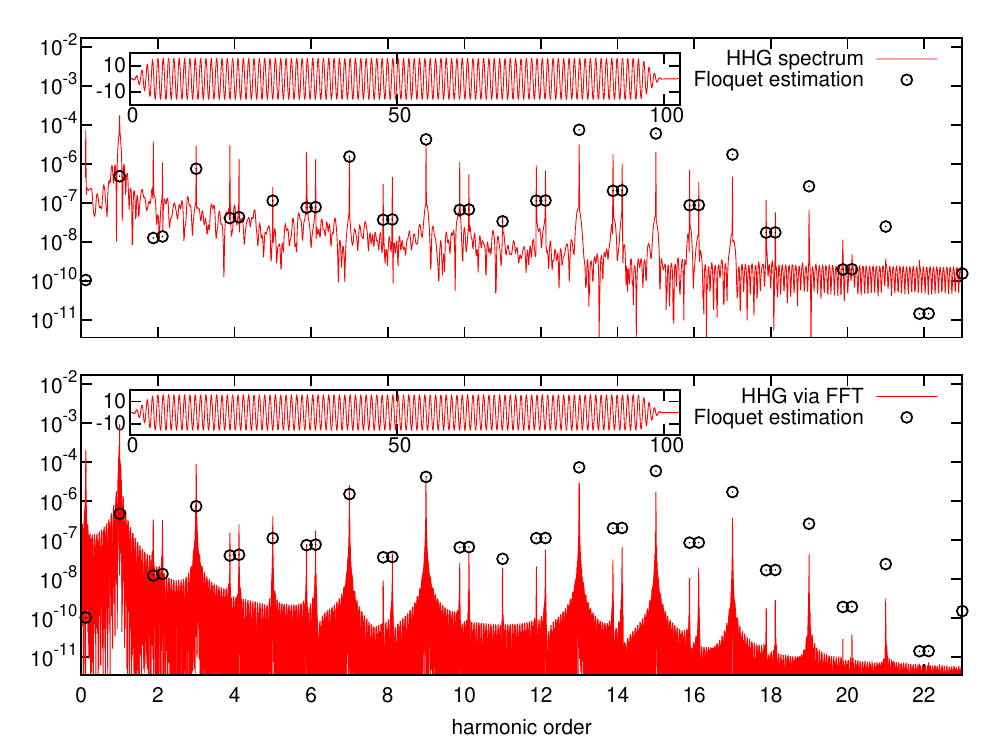}
\caption{HHG spectra obtained in different ways. Open circles on both panels correspond to the Floquet estimation. (The parameters are the ones that were denoted by the white circle in Fig.~\ref{Floquetfig}.) The finite duration pulse that mimics the corresponding monochromatic excitation is shown by the insets (where time on the horizontal axis has units of $T$, and we plotted $dE(t)/\hbar\omega_0,$ similarly to Figs.~\ref{nexpfig1} and \ref{nexpfig2}). The spectra plotted by solid red lines were obtained using our model based on the expectation values (top panel) and the power spectrum of $d^2 \langle D \rangle/dt^2$ (bottom panel). We normalized the results so that they have the same value at the strong, single peak at $\tilde{\omega}/\nu=9.$}%
\label{spectra2fig}%
\end{figure}

The finite width of the HHG peaks in Fig.~\ref{resnonresfig} is due to the finite duration of the exciting pulse, as well as to the fact that the envelope of the exciting pulse is not constant. (See Fig.~\ref{Floquetfig} for the sensitivity of $\delta\epsilon$ on $E_0.$) So in order to be able to perform a quantitative comparison between the predictions of the Floquet analysis and the numerical results, an exciting pulse with an envelope that is constant on a relatively long time interval is needed. To this end, let us relax the waveform (\ref{field}) and consider a simple sinusoidal excitation the duration of which is $100$ optical cycles, so that $E(t)$ reaches its maximum during the first $5$ cycles, and decays to zero during the last $5$ cycles (see the insets in Fig.~\ref{spectra2fig}). We compare a representative HHG spectrum obtained using the model of the previous section with results of the Floquet analysis and the Fourier transform based power spectrum of $d^2 \langle D \rangle/dt^2$ (which is usually considered as the source of the secondary radiation \cite{B09}) in Fig.~\ref{spectra2fig}.  As we can see, the positions of the peaks are practically the same for the Floquet estimation and the numerical result, and even the relative heights of the peaks are qualitatively similar. Additionally, there are remarkable similarities between the spectra that are based on the photon number expectation values and the Fourier transform of $d^2 \langle D \rangle/dt^2.$

Finally, let us return to Fig.~\ref{resnonresfig}. As we can see, for "blue-detuned" excitation (i.e., when $\nu>\omega_0$), there are considerably less HHG peaks that are visible than for the resonant case, and the number of these peaks is the highest for red detuning ($\nu<\omega_0$). This is a consequence of the general fact that $\frac{E_0 d}{\hbar\nu}$ is the parameter that determines the number of the relevant high harmonics in the Floquet time evolution (as shown by a simple transformation into dimensionless units, see e.g.~\cite{SBCF13}). That is, if all other parameters are the same, an excitation with lower frequency produces higher number of observable harmonics. As a remarkable example, see Fig.~\ref{resnonresfig}(d), where $\nu=\omega_0/10.$ This case of strong red detuning (that is relevant for real atoms and pulses in the infrared) is interesting also from the viewpoint of short pulse generation: the quasi-continuum of the frequencies corresponding to the plateau can be shown to have phases that produce short bursts of radiation in every half-cycle. Assuming $T=2\pi/\nu$ to have the order of magnitude of fs, the duration of these bursts (intensity fwhm) is around a few times 10 as.

\section{Photon statistics and correlations}
\label{schsec}
Besides the photon number expectation value, higher momenta as $\langle N_n^2\rangle$ are also required to determine the photon statistics of the HHG modes. However, the method introduced in Sec.~\ref{expdynsec} cannot be extended in a simple way, due to the increasingly complex hierarchy of the Heisenberg equations.

The physical assumptions that led to Eqs.~(\ref{a1})-(\ref{a10}) offer an alternative approach to obtain photon statistics. Since even the collection of all the modes that are taken into account affects negligibly the atomic dynamics on the timescale of the HHG, and there is no direct mode-mode interaction, the time evolution of a given mode is not strongly perturbed by the presence of the other ones. Thus it is reasonable to neglect all but one modes, which formally means reducing the sum in Eq.~(\ref{Hint}) to a single term. For the one and only mode that we take into account, the calculation of the photon statistics means no technical difficulty, the time dependent Schr\"{o}dinger equation can be solved e.g., in Schr\"{o}dinger picture by expanding the atom-mode wave function $|\Psi_{am}\rangle$ in the finite basis of $\left\{ |n,e\rangle, |n,g\rangle\ n=0,\ldots M\right \},$ where $M$ is the maximal photon number we take into account. The dynamics turns out to be independent from the truncation of the single mode Fock space already for $M\approx 10.$

\begin{figure}[htb]
\includegraphics[width=8.5cm]{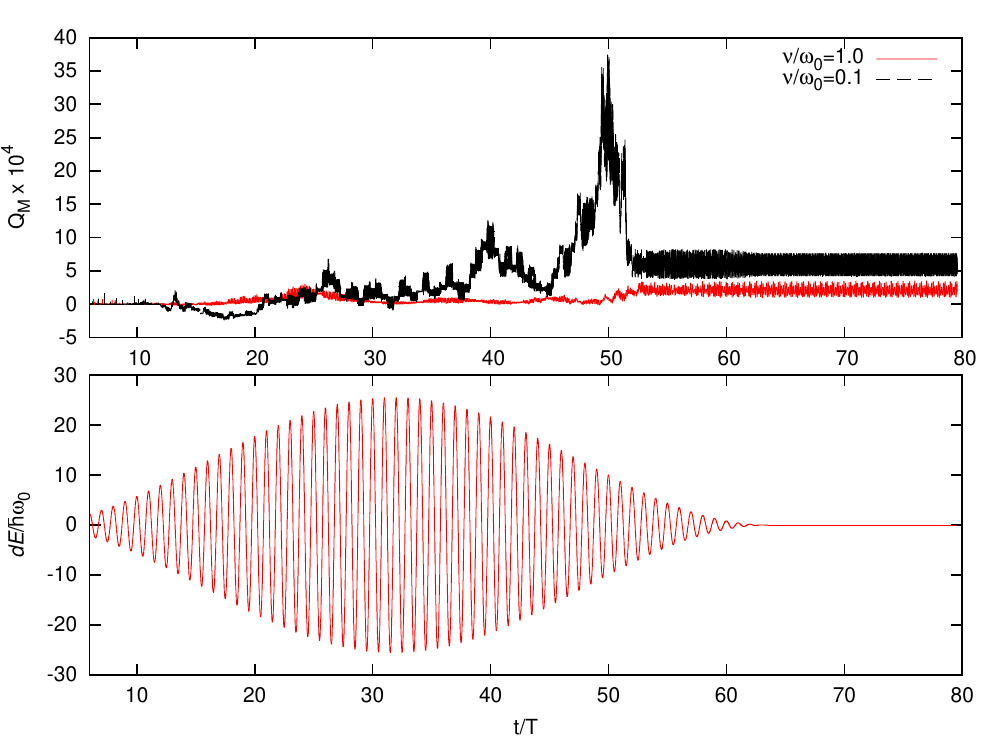}
\caption{The Mandel parameter (\ref{QM}) and the exciting classical pulse as a function of time, for the eights harmonics, ($\tilde{\omega}/\nu=8)$ both in the resonant, and strongly detuned cases.}%
\label{mandelfig}%
\end{figure}

The most general result (which is independently of the frequency $\tilde{\omega}$ of the mode) is that the probabilities
\begin{equation}
P_n(t)= \left|\langle n,e|\Psi_{am}(t) \rangle \right|^2 + \left|\langle n,g|\Psi_{am}(t) \rangle \right|^2
\end{equation}
are fast decreasing functions of the photon number $n.$ In agreement with the fact that the photon number expectation values are much below unity, $P_0$ dominates the photon statistics by being orders of magnitude larger than probabilities that correspond to nonzero photon numbers. Thus the HHG modes are very close to the vacuum, and there can be time instants when the photon number expectation value becomes (exactly or numerically) zero.  This circumstance causes difficulties when we are interested in e.g., the Mandel-parameter
\begin{equation}
Q_M=\frac {(\Delta N)^2}{\langle N \rangle}-1.
\label{QM}
\end{equation}
However, with careful evaluation, we see that $Q_M$ is typically a small (around $10^{-4}$), positive number, indicating super-Poissonian statistics. In the presence of the exciting field, there can be short intervals, in which $Q_M$ is negative, but finally, when the exciting field is zero again, $Q_M$ became positive for all the cases we investigated. Representative examples are shown in Fig.~\ref{mandelfig}. The narrow peaks at the early stage of the time evolution are signatures of the initial transient effects (see Figs.~\ref{nexpfig1} and \ref{nexpfig2}). Then rapid oscillations appear, which have considerably more regular pattern when the pulse is over ($t>\tau$). These "final" oscillations are solely due to the interaction of the atomic system and the quantized mode. (One may consider it as a consequence of a very far detuned, tiny amplitude Rabi flopping.) Note that when $\omega_0\gg \nu,$ the Mandel-parameter have larger values.

Numerically, for all harmonics up to 30 (and for the same exciting pulse as in Fig.~\ref{mandelfig}), we observed that the minimal value of $Q_M$ is between $-4\times 10^{-4}$ and $+ 10^{-5},$ while the maxima are between $10^{-4}$ and $0.04.$ During the oscillations for $t>\tau,$ $Q_M$ was always positive, with its time average being between $10^{-6}$ and $0.03.$ For excitations with different waveforms, the order of magnitude of these values were the same.

\bigskip

Since the maximal photon number that we need to take into account in order to obtain reliable results ($M$) is not too large, it is also possible to consider the problem with two quantized modes. (The corresponding Hilbert space is $2\times M^2$ dimensional.) The frequencies $\tilde{\omega}_1$ and $\tilde{\omega}_2$ of these modes can be chosen arbitrarily, which allows us to systematically check the validity of the assumptions that led to Eqs.~(\ref{a1})-(\ref{a10}). By calculating expectation values of the form $\langle A a_i a_j\rangle,$ $\langle A a_i^\dagger a_j\rangle$ and $\langle A a_i^\dagger a_j^\dagger\rangle$ (where $A=\tilde{U},$ $\tilde{V}$ or $\tilde {W}$, see the Appendix), it turns out that for the initial conditions (\ref{vac}), they mean negligible contributions to the dynamical equations. Numerically, the terms we neglected in Eqs.~(\ref{a1})-(\ref{a10}) are four orders of magnitude smaller than the ones we kept. The terms $\langle A a_i^\dagger a_i\rangle=\langle A N_i\rangle$ become important when $\langle N_i\rangle$ has the order of magnitude of unity at $t=0,$ and in this case the factorization $\langle A N_i\rangle=\langle A \rangle  \langle N_i\rangle$ is an accurate approximation.

Besides verifying former approximations, the model with two modes can also be used for the calculation of mode-mode cross correlations. In the Schr\"{o}dinger picture, the equal time second order correlation function
\begin{equation}
g_{ij}^2(t,t)=\frac{\langle \Psi(t)|N_i N_j |\Psi(t)\rangle}{\langle \Psi(t)|N_i|\Psi(t)\rangle \langle \Psi(t)|N_j |\Psi(t)\rangle}
\label{g2}
\end{equation}
can be -- in principle -- calculated straightforwardly. The general result is that regardless whether $i=j$ or not, $g^2_{ij}$ is larger than unity. (Note that the former case is in accord with the slightly super-Poissonian statistics we discussed above.) However, a systematic analysis of these correlations requires sophisticated numerical methods which are under development. These difficulties suggest that a transparent, analytically solvable model -- which is obviously not easy to construct -- would be very useful for the deep understanding of the process of HHG.

\section{Conclusions}
\label{conclusionsec}
We introduced a quantum optical model for the process of high harmonic generation (HHG) in a simple model system, a two-level atom. Considering a large number of electromagnetic modes with frequencies ranging from practically zero to the 30th harmonics of the classical, pulsed exciting field, we analyzed the time evolution of the photon number expectation values. We observed that high harmonics appear one after the other, modes with higher frequencies become visible later, when the envelope of the excitation is large enough to populate these modes. The final distribution of the photon number operators (when the exciting pulse is over) was shown to be close to traditional, Fourier transform based HHG spectra. For long enough pulses, the fine structure of the spectral peaks were shown to be closely related to the corresponding Floquet quasi-energies and eigenstates. The low photon number expectation values allowed us to analyze the single mode photon statistics of the high harmonic modes, and we found that these states have slightly super-Poissonian statistics, i.e., they are states with positive Mandel parameter $Q_M.$

Although most of these results are specific to the two-level systems, some of the methods we applied, and also a few of the physical consequences, can be transferred to more complex radiating systems, atoms and molecules as well. However, solving the coupled dynamical equations for a radiating system and a large number of electromagnetic modes (in a way analogous to the method presented in Sec.~\ref{modelsec}), can cause technical difficulties already for a hydrogen atom. On the other hand, the coupling of a single, quantized high harmonic mode is generally feasible. This would allow the investigation of the photon statistics also for more realistic systems.

\begin{acknowledgements}
Our work was supported by the Hungarian National Research, Development and Innovation Office under Contracts No. 81364 and K16-120615.
Partial support by the ELI-ALPS project is also acknowledged. The ELI-ALPS project (GOP-1.1.1-12/B-2012-000, GINOP-2.3.6-15-2015-00001) is supported by the European Union and co-financed by the European Regional Development Fund.
\end{acknowledgements}

\section{Appendix \\ (Equations of motion for the expectation values)}
For the consistency of the notation, let us use $U=\langle \sigma_x \rangle=\langle \hat{U} \rangle,$ $V=\langle \sigma_y \rangle=\langle \hat{V} \rangle$, $\ldots$ (i.e., characters without hat denote the expectation values of the corresponding operators decorated with hat, see Sec.~\ref{expdynsec}).   Eqs.~(\ref{der1})-(\ref{der3}) and (\ref{der5}) can be reformulated easily using $\hat{U}^\pm_n, V^\pm_n$ and $W^\pm_n.$ The exact Heisenberg equations of motion for these operators read:
\begin{eqnarray*}
\dfrac{d}{d t} \hat{U}_n^+ &=& \omega_0 \hat{V}_n^+ - \tilde{\omega}_n \hat{U}_n^-,\\
\dfrac{d}{d t} \hat{U}_n^- &=& \omega_0 \hat{V}_n^-
+ \tilde{\omega}_n \hat{U}_n^+ + \Omega_n, \\
\dfrac{d}{d t} \hat{V}_n^+ &=& -\omega_0 \hat{U}_n^+ - \tilde{\omega}_n \hat{V}_n^- + \Omega(t)\hat{W}_n^+ +\\
&+&\hat{W} \sum_{j}\Omega_j \left(a_n + a_n^\dagger\right)\left(a_j + a_j^\dagger\right),\\
\dfrac{d}{d t} \hat{V}_n^- &=& -\omega_0 \hat{U}_n^-
+ \tilde{\omega}_n \hat{V}_n^+
+ \Omega(t) \hat{W}_n^- +\\
&+&i\hat{W}\sum_{j}\dfrac{\Omega_j}{2}\left\{\left(a_n - a_n^\dagger\right),\left(a_j + a_j^\dagger\right)\right\},\\
 \dfrac{d}{d t} \hat{W}_n^+ &=& -\tilde{\omega}_n \hat{W}_n^- - \Omega(t)\hat{V}_n^+ -\\
&-& \hat{V} \sum_{j}\Omega_j \left(a_n + a_n^\dagger\right)\left(a_j + a_j^\dagger\right),\\
\dfrac{d}{d t} \hat{W}_n^- &=& \tilde{\omega}_n \hat{W}_n^+
- \Omega(t) \hat{V}_n^- -\\
&-&i\hat{V} \sum_{j}\dfrac{\Omega_j}{2}\left\{\left(a_n - a_n^\dagger\right),\left(a_j + a_j^\dagger\right)\right\},
\end{eqnarray*}
where $\{.,.\}$ stands for the anticommutator. Approximations take place when we calculate the expectation values of both sides of the equations above. As it has been mentioned in
Sec.~\ref{schsec}, for any atomic operator $A,$ it is consistent to set $\left\langle A a_n a_j\right\rangle=\left\langle A a_n^\dagger a_j^\dagger\right\rangle=0,$ and to consider terms $\left\langle A a_n a_j^\dagger\right\rangle,$ $\left\langle A a_n^\dagger a_j\right\rangle$ to be nonzero only for $n=j.$ This, with factorizations $\left\langle A a_n a_n^\dagger\right\rangle=\left\langle A\right \rangle \left \langle  a_n a_n^\dagger\right\rangle,$ $\left\langle A a_n^\dagger a_n\right\rangle=\left\langle A\right \rangle \left \langle  a_n^\dagger a_n\right\rangle,$ result in:
\begin{eqnarray}
\dfrac{d}{d t} U_n^+ &=& \omega_0 V_n^+ - \tilde{\omega}_n U_n^-\label{a1},\\
\dfrac{d}{d t} U_n^- &=& \omega_0 V_n^-
+ \tilde{\omega}_n U_n^+ + \Omega_n, \\
\dfrac{d}{d t} V_n^+ &=& -\omega_0 U_n^+ - \tilde{\omega}_n V_n^- + \Omega(t)W_n^+ + \nonumber \\
&+&W \Omega_n \left(2\left\langle N_n \right\rangle+1\right),\\
\dfrac{d}{d t} V_n^- &=&-\omega_0 U_n^-
+ \tilde{\omega}_n V_n^+
+ \Omega(t) W_n^-,\\
 \dfrac{d}{d t} W_n^+ &=& -\tilde{\omega}_n W_n^- - \Omega(t)V_n^+ -V \Omega_n \left(2\left\langle N_n \right\rangle+1\right),\\
\dfrac{d}{d t} W_n^- &=& \tilde{\omega}_n W_n^+
- \Omega(t) V_n^-.
\end{eqnarray}
For the sake of completeness, let us express the expectation values of Eqs.~(\ref{der1})-(\ref{der3}) and (\ref{der5}) using the dynamical variables:
\begin{eqnarray}
\frac{d}{dt}U&=&\omega_0 V,\\
\frac{d}{dt}V&=&-\omega_0 U + \Omega(t) W +\sum_n \Omega_n W_n^+,\\
\frac{d}{dt}W&=&- \Omega(t) V -\sum_n \Omega_n V_n^+,\\
\frac{d}{dt}\langle N_n \rangle &=&\frac{\Omega_n}{2}U_n^-. \label{a10}
\end{eqnarray}

\bibliography{allthebib}

\end{document}